\documentclass{kluwer}    
\usepackage{psfig}
\newdisplay{guess}{Conjecture}

\begin{document}                                                                                   
\begin{article}
\begin{opening}         
\title{Identification of Shocks in the Spectra from Black Holes}
\author{Samir \surname{Mandal}\email{space\_phys@vsnl.com}}  
\institute{Centre for Space Physics, Chalantika 43, Garia Station Rd., Kolkata 700084}
\author{Sandip Kumar \surname{Chakrabarti}\email{chakraba@bose.res.in}}  
\runningauthor{S. Mandal and S.K. Chakrabarti}
\runningtitle{Identification of Shocks in Black Hole Spectra}
\institute{S.N. Bose National Centre for Basic Sciences, JD Block, Salt Lake, Kolkata 700098\\
and\\
Centre for Space Physics, Chalantika 43, Garia Station Rd., Kolkata 700084\\ }
\date{August 26th, 2004}

\begin{abstract}
We study the spectral properties of a low angular momentum flow
as a function of the shock strength, compression ratio,
accretion rate and flow geometry. In the absence of a satisfactory
description of magnetic fields inside the advective disk, we
consider the presence of only stochastic fields and use the
ratio of the field energy to the gravitational energy density as a 
parameter. We not only include `conventional' synchrotron emission 
and Comptonization by Maxwell-Bolzmann electrons in the gas, but
we also compute these effects due to power-law electrons. 
For strong shocks, a bump is produced due to the post-shock flow.
A power-law spectral components due to the thermal and non-thermal electrons appear after this
bump.
\end{abstract}


\end{opening}           

\noindent To appear in Astrphysics and Space Sciecne (Proceeding of the Hong Kong Conference; edited 
by K.S. Cheng and Romero)

\section{Introduction}  

In Chakrabarti \& Titarchuk (1995, hereafter CT95), a two component advective
flow (TCAF) solution was introduced (see, Chakrabarti, this volume). The sub-Keplerian 
flow has a steady shock but depending on the flow parameters it could 
be oscillating or propagating. The post-shock region is called CENtrifugal pressure supported
BOundary Layer or CENBOL. Given that the accretion flow largely passes through this shock,
it is likely that the hot electrons may be {\it accelerated} by the shock just as
high energy cosmic rays are produced by the transient super-novae shocks
(Bell 1978ab, Longair 1981). We assume this process in computing the spectra
and show that the presence of shocks make their mark distinctly. Our solution also 
produces gamma-rays naturally since shocks are generally produced due to the 
centrifugal barrier (e.g., Chakrabarti, 1996). It is possible that 
occasionally an additional source of high energy photons, perhaps in the form 
of shocks in outflows and jets, is necessary to explain the complete spectra.
Details of our model are presented in Mandal \& Chakrabarti (2004).

\section{Relevant heating and cooling processes and the solution procedure}

Protons loss energy through Coulomb interaction and inverse bremss -trahlung.
Electrons are heated through this Coulomb coupling. Electrons are cooled by
bremsstrahlung, cyclo-synchrotron and Comptonization of the soft photons
due to cyclo-synchrotron radiation. The Comptonization is computed by using 
this cooling term augmented by the enhancement factor. In the present situation, 
the flow passes through the shocks and thus the electrons will have a power-law distribution 
Thus, both the thermal and the non-thermal electrons are present. 

The geometry of the flow is chosen to be conical for simplicity and the flow surface
makes an angle $\Theta$ with the z-axis. We assumed the form of infall velocity $v(x) 
=v_{\rm 0} x^{-1/2}$. At the shock, $x=x_{\rm s}$, the velocity is reduced by a factor $R$, the
compression ratio, i.e., $v_+(x_{\rm s})=v_-(x_{\rm s})/R$. The number density also goes up by 
a factor $R$. 

As for the initial conditions, we fix the outer boundary at a large distance
(say, $10^6 r_{\rm g}$) and supply matter (both electrons and protons) with the
same temperature (say, $T_{\rm p}=T_{\rm e}=10^6$K). By integrating energy equation, we
obtained the temperature. The radiation emitted by the flow through bremsstrahlung
and synchrotron radiation are inverse Comptonized by the hot electrons in the flow.

A shock of compression ratio $R$ causes the formation of power-law 
electrons of slope (Bell, 1978ab), $p=(R+2)/(R-1)$. This power-law 
electrons produce a power-law synchrotron emission with an index 
(Longair, 1981) $q=(1-p)/2$. The power-law electrons have energy minimum at
$ {\cal E}_{\rm min} $ and energy maximum at $ {\cal E}_{\rm max} $. ${\cal E}_{\rm max}$ 
is obtained self-consistently by conserving the number of power-law electrons
and by computing the number of scatterings that the electrons undergo
inside the disk before they escape and ${\cal E}_{\rm min}$ is obtained
from the temperature of the injected electrons. This temperature
is obtained self-consistently through our integration procedure. 

In presence of magnetic field (assumed to be the ratio of the magnetic energy density 
to gravitational energy density), the pre-shock flow and the CENBOL would emit
synchrotron radiation. We assume that the number of black body photons 
generated by the electrons obeying power-law distribution (assumed to be 
of constant fraction $\zeta$ of thermal electrons) below self-absorption frequency
is the same fraction $\zeta$ of the total black body photons. 
We followed the procedures presented in CT95
and Titarchuk \& Lyubarskij (1995) while computing the Comptonized spectrum
due to scattering from thermal electrons and power-law electrons. However, unlike CT95
where a Keplerian  disk was supplying soft photons, here the source is
everywhere, i.e., distributed throughout. This has been taken into account.
At the end, we add contributions from all the components to get
the net photon emission from the flow.

In the next Section, we present the spectral characteristics
by varying parameters such as $\Theta$, $x_{\rm s}$, $R$, $\zeta$ and ${\dot m}$.
We use a  black hole of mass $10M_\odot$.

\section{Results and interpretations}

In Fig. 1 we show our simplified flow geometry to capture the essential physics
and show how the temperature of the electrons and protons differ
as the flow moves closer to the black hole. The deviation is more prominent 
in the post-shock region. Here, the radial distance $x$ is measured 
in units of the Schwarzschild radius $ r_{\rm g} $.

\begin {figure}
\vbox{
\vskip -2.0cm
\hskip 0.5cm
\centerline{
\psfig{figure=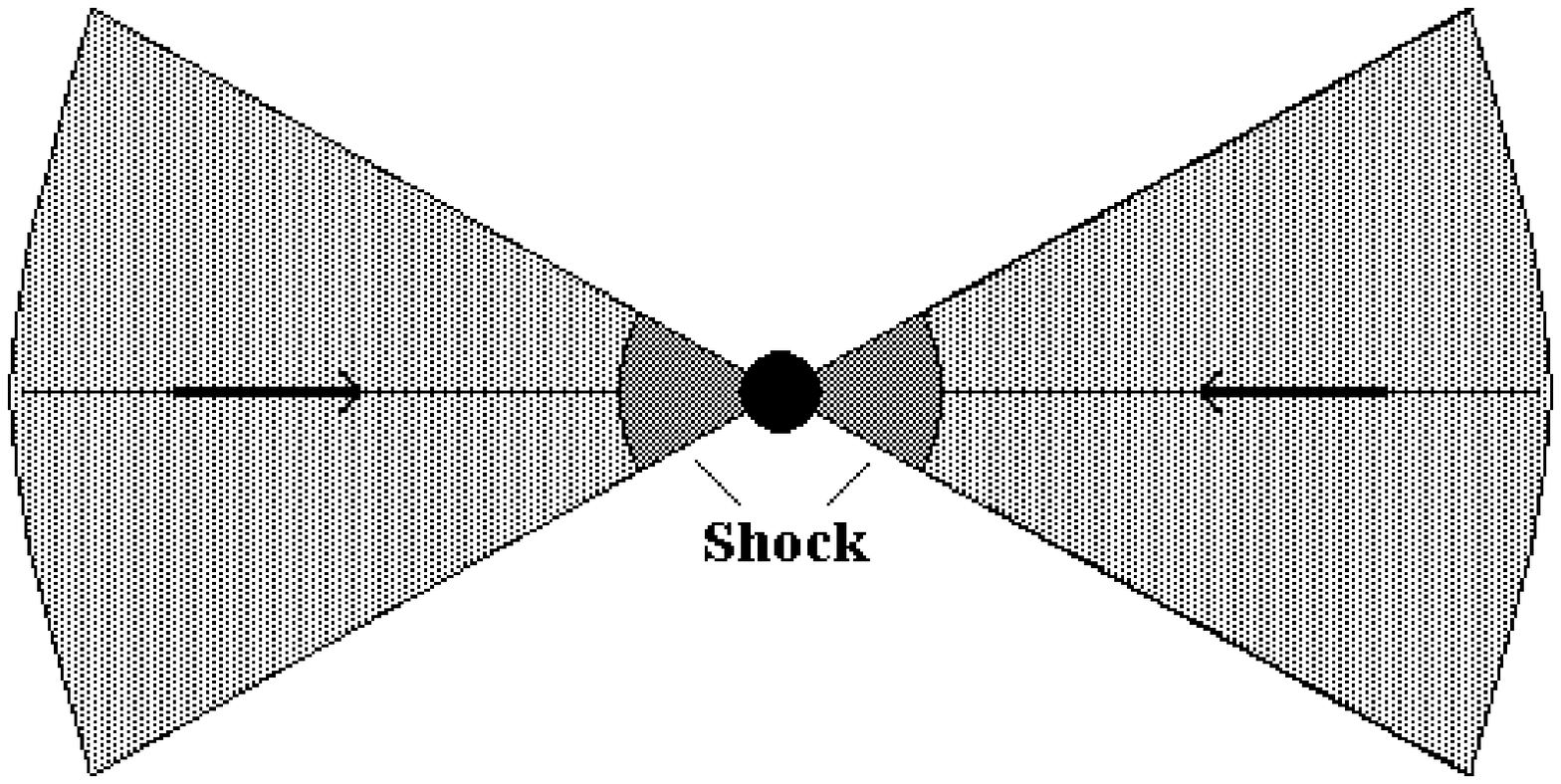,height=5truecm,width=5truecm}
\psfig{figure=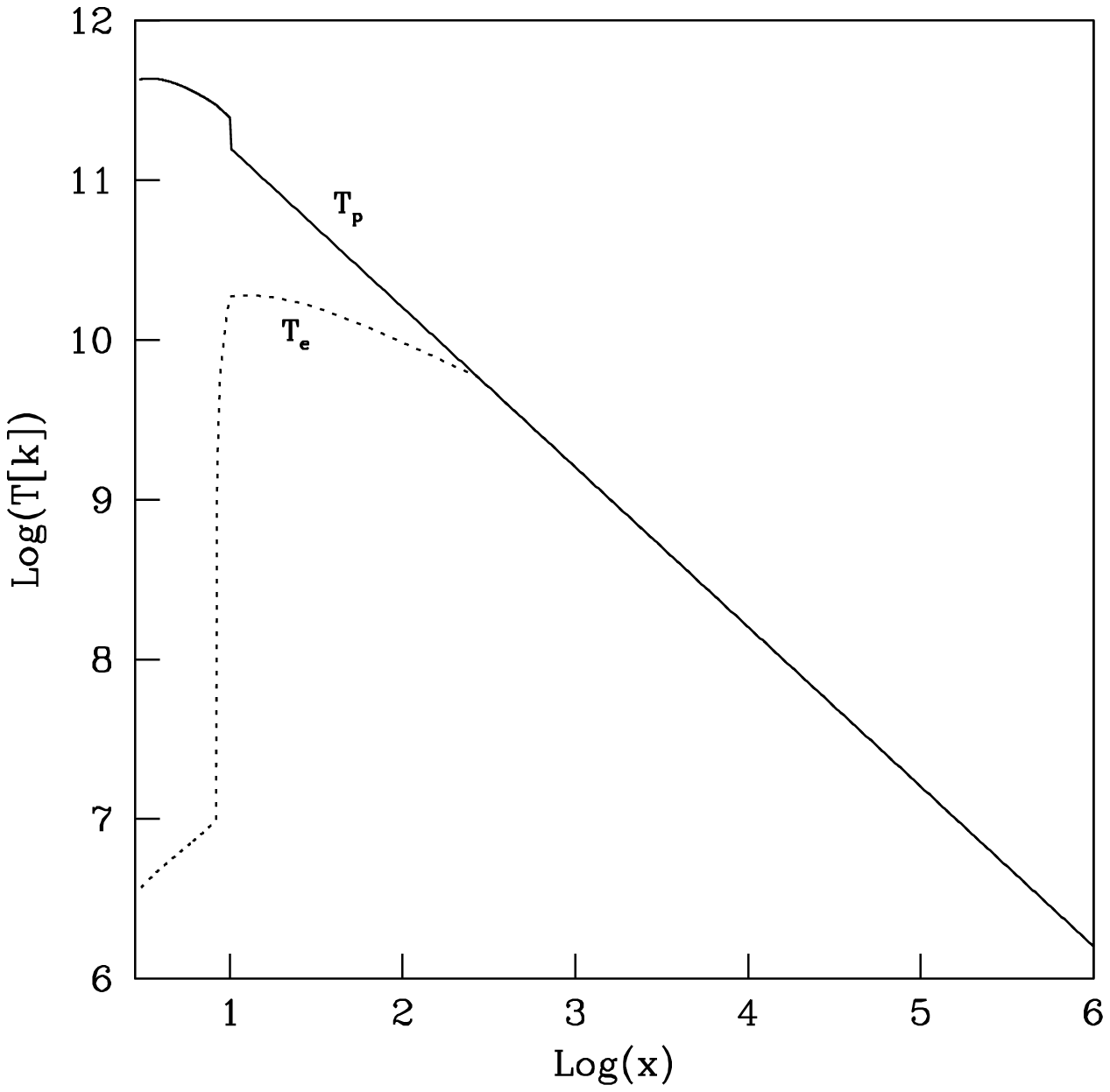,height=8truecm,width=8truecm}}}
\vspace{0.0cm}
\caption{(left) Wedge shaped flow model assumed in this paper which 
included a shock at the inner region. (right)
Variation of proton and electron temperatures (marked as $T_{\rm p}$ and $T_{\rm e}$) as
functions of dimensionless radial distance $x$ for ${\dot m}=0.00059$.
Other parameters are $M=10M_\odot$, $R=3.9$, $\Theta=77^o$, $\zeta=0.7$
and $x_{\rm_s}=10$. Closer to the black hole, the electrons
cool more rapidly than the protons.}
\end{figure}

In Fig. 2, we show a typical spectrum obtained from our solution. 
The curve marked 1 is due to pre-shock flow and the curve marked 2 is due 
to the post-shock flow. The synchrotron self-absorption frequency at $\nu_{\rm a}$
is marked. The bump due to post-shock flow is at frequency $\nu_{\rm bump} \sim \nu_{\rm inj} [1+\frac{4}{3}\frac{R-1}
{R}\frac{1}{x_{rm s}^{1/2}}]^{x_{\rm s}^{1/2}}$, where, $\nu_{\rm inj}$ is the
frequency of the dominant photons from the pre-shock flow, $R$
is the compression ratio of the shock located at $x_{\rm s}$. The power law is 
produced by the power-law electrons from the shock acceleration.

\begin{figure}
\vbox{
\vskip -3.0cm
\hskip 0.0cm
\centerline{
\psfig{figure=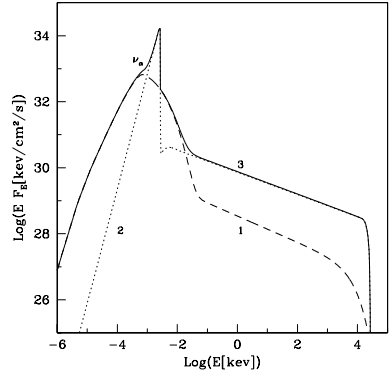,height=8truecm,width=8truecm}}}
\vspace{0.0cm}
\caption[]{ 
A typical complete spectrum from a sub-Keplerian accretion flow with all
the contributions shown for the same parameters as in Fig. 1b.
The curve marked `1' is due to Comptonization of the synchrotron emission from the
pre-shock region.  The curve marked `2' is due to the Comptonized spectra from the
post-shock (CENBOL) region.  The curve marked `3' indicates the net
radiation from the flow. $\nu_{\rm a}$ represents synchrotron self-absorption frequency for the CENBOL.
}
\end{figure}

In Fig. 3, we compare three cases with shock locations at 
$x_{\rm s}=10, \ 40$ and $80$ respectively. The shock locations can vary 
with specific angular momentum in the flow which in turn is governed
by the viscosity parameter (Chakrabarti, 1990; 1996).
\begin{figure}
\vbox{
\vskip -3.0cm
\hskip 0.0cm
\centerline{
\psfig{figure=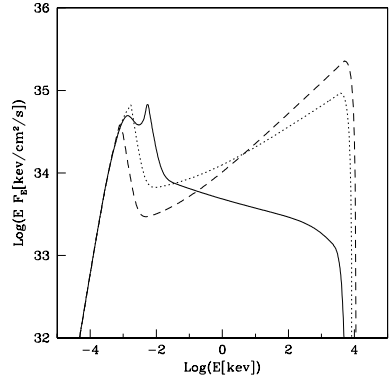,height=8truecm,width=8truecm}}}
\vspace{0.0cm}
\caption[]{
Variation of the total emitted spectrum when the shock
location $x_{\rm s}$ is varied. The solid, dotted and dashed curves are drawn
for $x_{\rm s}=10,\ 40,\ 80$ respectively. The spectrum becomes softer when
 $x_{\rm s}$ is reduced. Here, ${\dot m }=0.1$ has been chosen.
}
\end{figure}

With the rise in the accretion rate of the
sub-Keplerian disk, the spectrum is likely to become harder. This can be seen more
clearly in Fig. 4, where we vary the accretion rate. A higher accretion rate is
expected to increase the density of the flow and it becomes difficult to cool the matter
by the soft photons of the synchrotron radiation. The Comptonized spectrum becomes harder.
At lower accretion rate, it is easy to cool the flow and the spectrum becomes softer.
The curves are drawn for: ${\dot m}=0.05$ (solid), $0.5$ (dotted) and $1.5$ (dashed).

\begin{figure}
\vbox{
\vskip -3.0cm
\hskip 0.0cm
\centerline{
\psfig{figure=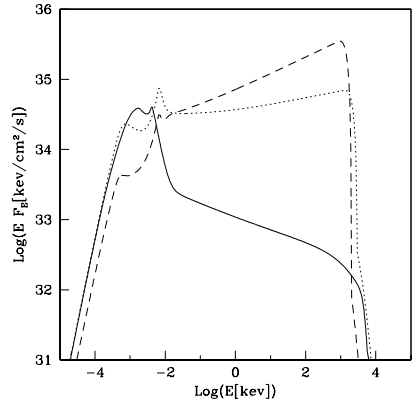,height=8.5truecm,width=8.5truecm}}}
\vspace{0.0cm}
\caption[]{Variation of the spectra with the dimensionless accretion rate. Solid,
dotted and dashed curved are for ${\dot m}=0.05,\ 0.5\ and\ 1.5$
respectively. The spectrum becomes harder when the accretion rate is
increased.}
\end{figure}

\begin{figure}
\vbox{
\vskip -3.0cm
\hskip 0.0cm
\centerline{
\psfig{figure=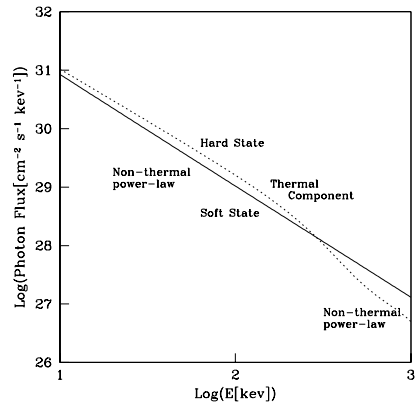,height=8.5truecm,width=8.5truecm}}}
\vspace{0.0cm}
\caption[]{
Example of a typical soft state  and a hard state spectra of a galactic black hole candidate
os mass $10M_\odot$. The parameters for the hard state are: 
${\dot m}=2.4$, $R=4$, $x_{\rm s}=250$, $\Theta=80$, $\zeta = 0.24$. 
The parameters for the soft state are:
${\dot m}=0.02$, $R=4$, $x_{\rm s}=100$, $\Theta=80$, $\zeta = 0.3$.
}
\end{figure}

\section{A typical soft/hard spectra in gamma-ray region}

Typical broad band soft state and hard state spectra are presented in Fig. 4 of 
Chakrabarti (this volume). In Fig. 5 of Chakrabarti (this volume) behaviour at 
high energy has been shown for Cyg X-1, GROJ1719-24 and GROJ0422+32 (taken from 
Ling \& Wheaton, 2004). In Fig. 5 we present the photon flux in soft and hard
states as obtained by our solution with shocks. No jet contribution was added. 
Not surprisingly, they look similar. However, a second kind of high $\gamma$ intensity state was reported 
(see, Case et al. 2004) in which non-thermal power-law has been observed from a few 
tens of keV to few MeVs (see, Fig. 4 of Chakrabarti, this volume). 
It has been difficult to obtain such a behaviour without invoking jets or other mechanisms.

\acknowledgements

SM acknowledges a RESPOND project from Indian Space Research Organization which supports his research.

\end{article}
\end{document}